\documentclass[12pt]{article}

\usepackage{amsfonts}
\usepackage{amssymb}
\usepackage{verbatim}
\usepackage{cleveref}
\usepackage{graphicx}
\usepackage{enumerate}
\usepackage{color}
\usepackage{url}

\usepackage{enumitem}
\setlist[itemize]{noitemsep, topsep=0pt}

\usepackage[ruled,vlined,linesnumbered]{algorithm2e}
\Crefname{algocf}{Algorithm}{Algorithms}
\SetKw{Continue}{continue}
\SetKw{Break}{break}

\newcommand*{\QED}{\hfill\ensuremath{\square}\vspace{0.5em}}%
\newcommand\pto{\mathrel{\ooalign{\hfil$\mapstochar\mkern5mu$\hfil\cr$\to$\cr}}}

\newtheorem{definition}{Definition}[section]

\newtheorem{proposition}{Proposition}[section]
\newtheorem{corollary}{Corollary}[section]
\newtheorem{theorem}{Theorem}[section]

\title{The compressions of reticulation-visible networks are tree-child}
\author{Andreas D.M. Gunawan,  Hongwei Yan,  Louxin Zhang\thanks{to which correspondence should be addressed. email: matzlx@nus.edu.sg.  Authors' address: Department of Mathematics, National University of Singapore, 
Singapore 119076.}}
\date{}

\begin{document}

\maketitle

\begin{abstract}
 Rooted phylogenetic networks are rooted  acyclic digraphs. They are used
to model complex evolution where hybridization, recombination and other reticulation events play important roles.
A rigorous definition of network compression is introduced on the basis of the recent studies of the relationships between cluster, tree and rooted phylogenetic network. The concept reveals another interesting connection between the two well-studied network classes\textemdash tree-child networks and reticulation-visible networks\textemdash and enables us to define a new class of networks for which the cluster containment problem has a linear-time algorithm.

\end{abstract}

\section{Introduction}

Genomes are  shaped not only by random mutations over  generations but also horizontal genetic transfers between individuals of
different species \cite{Doolittle_07,Marcussen_12}. Because of their high complexity, the evolutionary history of extant genomes are extremely hard to reconstruct. This motivates researchers to adopt rooted phylogeny networks (RPNs) to model both vertical and reticulate evolutionary events in comparative genomics.  The combinatorial and algorithmic aspects of RPNs have extensively been investigated in the last two decades \cite{Gusfield_book,Huson_Book_11,Steel_book,Wang_JCB_01}. 

One important issue of the inference of RPNs is how to define a good distance metric for  model verification. Phylogenetic trees are special RPNs for which the so-called Robinson-Foulds distance serves the purpose well. The Robinson-Foulds distance between two phylogenetic trees is defined to be the number of clusters appearing in one but not in the other.  However, defining the concept of cluster for RPNs is not straightforward, and  neither is computing the clusters displayed by an RPN \cite{Huson_Book_11,Moret_04,Zhang_18}. In the same spirit, a tree-based distance metric is also introduced for RPNs \cite{Huson_Book_11}.  This is one of the reasons for the introduction of tree-child networks~\cite{Cardona_TCBB_09}, galled networks~\cite{Huson_07}, tree-based networks~\cite{Francis_15}, etc.  The combinatorial characterizations of  these networks and connections between these networks are two hot topics of the current research in the phylogenetic network community.  

Consider a phylogenetic tree and an RPN  that are over the same set of taxa.  The RPN displays the tree if the network contains as a spanning tree some subdivision of the tree. 
The tree containment problem is determining whether or not $N$ displays $T$ given an RPN $N$ and a phylogenetic tree $T$. The cluster containment problem is determining whether $S$ is a cluster in some tree displayed by $N$ given an RPN $N$ over $X$ and $S\subseteq X$ \cite{Kanji_08,Nakhleh_05,Iersel_10}.   In the study of these two  algorithmic problem,  simplifying the input RPN by  replacing  a set of connected non-leaf nodes with a single node of the same type turns out to be useful \cite{Yan_2018, Zhang_18}. This technique is also useful for bounding the number of the nodes of different types in an RPN \cite{Gunawan_IC_17} and other studies \cite{Cardona_TCBB_09,Huber_2016}.  Motivated by these facts, we define  rigorously a compression operation for RPNs. Interestingly, the definition reveals another interesting connection between reticulation-visible networks and tree-child networks. It also allows us to define a new network class for which the cluster containment remains linear-time solvable. 

The rest of this paper is divided into four sections. In Section~\ref{sec2}, we introduce some basic concepts and terminology that facilitate the study. These concepts include the types of network nodes and the classes of RPNs.  In Section~\ref{sec3}, we introduce a rigorous definition of compressing RPNs and study some elementary properties of network compression.  Importantly, we will prove that the compression of a reticulation-visible network is tree-child.    Section~\ref{sec4} defines a new class of RPNs, namely quasi-reticulation-visible network, as an expansion of the class of reticulation-visible networks.  Lastly, we make a couple of remarks to conclude the paper in Section~\ref{sec5}.

\section{Basic Notation}
\label{sec2}

\subsection{Phylogenetic networks}

   Let $X$ be a set of taxa.  An RPN over $X$ is a  rooted acyclic digraph such that:
        \begin{itemize}
            \item[(i)] every edge is directed away from the root that is of indegree 0 and outdegree at least 1;
       \item[(ii)] every node other than the root is of either indegree 1 or outdegree 1; and  
       \item[(iii)] all the nodes of indegree one and outdegree 0, called the {\it leaves}, are bijectively labeled with the taxa in $X$.
     \end{itemize}
 In an RPN, a non-leaf node is called a {\it reticulate node} if its indegree is at least two (and its outdegree is 1). It is called a {\it tree node} if it is the root or it is of indegree 1 and outdegree at least 2. It is called a {\it redundant node}\footnote{Traditionally, an RPN does not contain any redundant nodes.  Here, we introduce such a type of nodes for presenting a rigorous definition of the network compression.} if  it is of  indegree   1 and outdegree 1.   

Let $N$ be an RPN $N$ over $X$. We use  $\mathcal{V}(N)$,  $\mathcal{T}(N)$, $\mathcal{R}(N)$, $\mathcal{D}(N)$ and $\mathcal{L}(N)$ to denote the sets of nodes, tree nodes, reticulate nodes, redundant nodes and leaves, respectively.  Clearly, $\mathcal{V}(N)=\mathcal{T}(N) \cup \mathcal{R}(N)\cup\mathcal{D}(N) \cup \mathcal{L}(N)$. We also use $\mathcal{E}(N)$ to denote the edge set of $N$.

An RPN is {\it binary} if every reticulate node is of indegree 2 and every tree node  is of outdegree 2. We allow degree-2 nodes to appear in a binary RPN. 

\begin{figure}[t!]
            \centering
            \includegraphics[scale = 0.8]{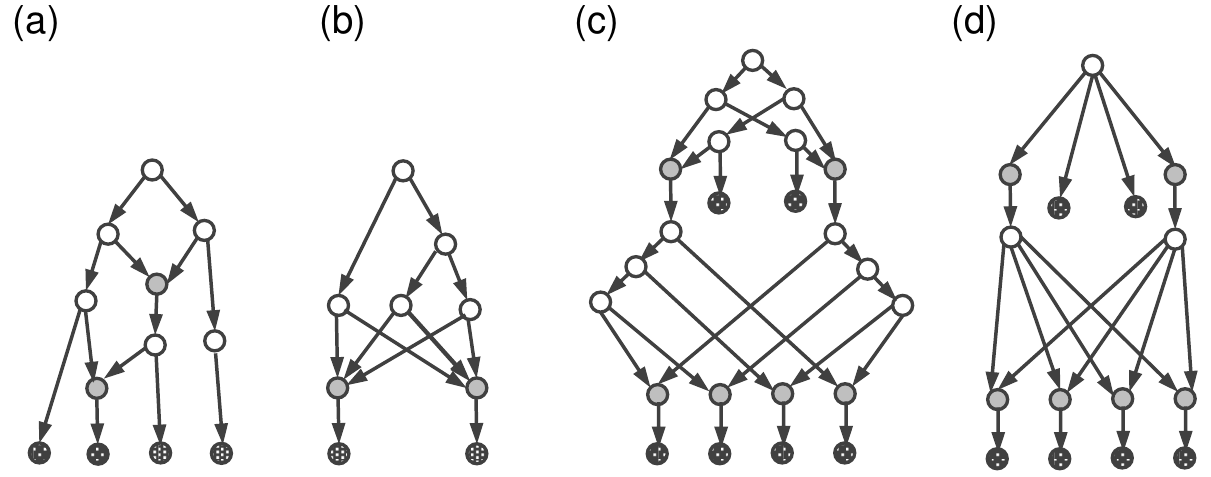}
            \caption{({\bf a}) A binary tree-child RPN in which there is a redundant node.  ({\bf b})  A non-binary RPN that is reticulation-visible, but not tree-based. ({\bf c}) A binary RPN that is tree-sibling, but not reticulation-visible. ({\bf d}) A non-binary RPN that is neither tree-sibling nor reticulation-visible. Here,  reticulate nodes and leaves are drawn as filled and pattern-filled  circles, respectively.  Leaf labels are omitted, as they are irrelevant in this study.}
            \label{Fig_0}
        \end{figure}
  
\subsection{Tree-based RPNs, tree-sibling RPNs and their subclasses}
       
 Let $N$ be an RPN.  For any $u, v\in \mathcal{V}(N)$,  we say that $u$ is a {\it parent} of $v$ and that  $v$ is the {\it child} of $u$ if $(u, v)\in \mathcal{E}(N)$.  
 A {\it directed path} from $u$ to $v$ is a sequence of nodes $u_0, u_1, \cdots, u_k$ ($k\geq 1$) such that $u_0=u$, $u_k=v$ and  
$u_i$ is a parent of $u_{i+1}$ for every $i\leq k-1$. 
The node $u$ is said to be an {\it ancestor} of $v$ if there is a directed path from $u$ to $v$.  We also say that $u$ is {\it above} $v$ if $u$ is an ancestor of $v$.  

Let $x$ and $y$ be two nodes of  $N$ such that $x$. The node $x$ is  a {\it dominator} of $y$ if (i) $x$ is an ancestor of $y$ and (ii) every directed path from the root of $N$ to $y$ contains $x$.  
%
 The node  $x$ is said to be \textit{visible} if it is a dominator of $\ell$ for some $\ell\in \mathcal{L}(N)$. 
        
   \begin{proposition}\label{prop_1}
            The following statements hold:
            \begin{itemize}
                \item[{\rm (1)}] A node is invisible if its children are all reticulate nodes.
                \item[{\rm (2)}] A node is visible if it has a tree or redundant child $y$ that is visible.
 \item[{\rm (3)}] If $u$ is a dominator of $v$ and $v$ is visible, then 
  $u$ is also visible.
                \item[{\rm (4)}]  A reticulate (resp. redundant) node is visible if and only if its unique child is a leaf,  a visible tree node or a visible redundant node.
            \end{itemize}
        \end{proposition}
{\bf Proof.}  Parts 1--2 are proved in \cite{Gambette_RECOMB_15}.  Part 3 follows from that the dominator relation is transitive \cite{Lengauer_79}.  Part 4 follows from Parts 1--3.
\QED
\vspace{0.5em}

The RPN $N$ is said to be {\it tree-child} if every non-leaf node $u$ has a child that is a leaf,  a tree node,  or a redundant node in $N$. It is easy to see that $N$ is tree-child if and only if, 
for any $u\in \mathcal{V}(N)$, a path from $u$ to a leaf exists such that every non-leaf node other than $u$ of the path is  either a tree node or a redundant node.

Note that every node is visible in $N$ if and only if $N$ is tree-child \cite[Proposition 3.1]{Gambette_RECOMB_15} (see also \cite{Steel_book,Zhang_18}). $N$ is said to be  {\it reticulation-visible}\footnote{This definition is identical to the current definition 
for RPNs without redundant nodes \cite{Huson_Book_11}.} if every reticulate or redundant node  is visible.  Hence, tree-child networks form a proper subclass of reticulation-visible networks. 

A network edge $(u, v)$ is called a reticulate edge if $v$ is a reticulate node. 
 $N$ is {\it tree-based} if there exists a set $E$ of $\sum_{v\in \mathcal{R}(N)}(d_o(v)-1)$ reticulate edges such that the subnetwork $(\mathcal{V}(N), \mathcal{E}(N) \backslash E)$ is a tree with the same set of leaves as $N$ \cite{Jetten_TCBB_18}, where $d_o(v)$ denotes the outdegree of $v$. Binary reticulation-visible networks are tree-based (\cite{Francis_15,Gambette_RECOMB_15}). However,  non-binary reticulation-visible networks may not be tree-based (Fig.~\ref{Fig_0}b) \cite{Jetten_TCBB_18}.

Two nodes are {\it sibling} if they have a common parent.  $N$ is said to be {\it tree-sibling} if every reticulate node has a sibling that is a tree node or a leaf. Note that tree-child RPNs form a subclass of tree-sibling RPNs.

\section{Network Compression}
\label{sec3}

\subsection{Decomposition of  RPNs}

Let $N$ be an RPN over a set of taxa. Recall that $\mathcal{T}(N)$ and $\mathcal{R}(N)$ denote the sets of tree nodes and reticulate nodes, respectively.  The subnetwork $N|_{\mathcal{T}(N)}$ induced by $\mathcal{T}(N)$ is a forest (shaded, Fig.~\ref{Fig_1}a). One connected component of the forest is a subtree rooted at the root of $N$, whereas the rest are each rooted at the child of some reticulation. These connected components are called the {\it tree-node components}\footnote{In the papers \cite{Gunawan_IC_17,Zhang_18} that were published earlier, each tree-node component is a connected component of the forest $N|_{\mathcal{T}(N)\cup \mathcal{L}(N)}$. Here, a tree-node component contains neither leaves nor redundant nodes.}. 

Similarly, the subnetwork $N|_{\mathcal{R}(N)}$ induced by $\mathcal{R}(N)$ is also a forest. Each connected component of this forest is called a {\it reticulation component} (circled by a dashed line, Fig.~\ref{Fig_1}a). It is easy to see that each reticulation component has a unique reticulate node such that any other node is above it. Such a node called the root of  the component. 

\begin{proposition}
 Let $N$ be an RPN over $X$ such that $\mathcal{D}(N)=\phi$ and $|X|=n$. Assume that $N$ contains $p$  tree-node components and $q$ reticulation components. Then, $p-1\leq q \leq n+p-1.$
\end{proposition}
{\bf Proof.} The first inequality follows from the fact that each tree-node component is rooted at either the root of $N$, $\rho(N)$, or the child of the lowest reticulate node of a reticulation component. 

The second inequality follows from that the lowest reticulate node of every reticulation component is the parent of a leaf or the root of a tree-node component  and that the tree node component rooted at $\rho(N)$ is not below any reticulate node.
\QED

\subsection{Compressing RPNs}

\begin{figure}[t!]
            \centering
            \includegraphics[scale = 0.8]{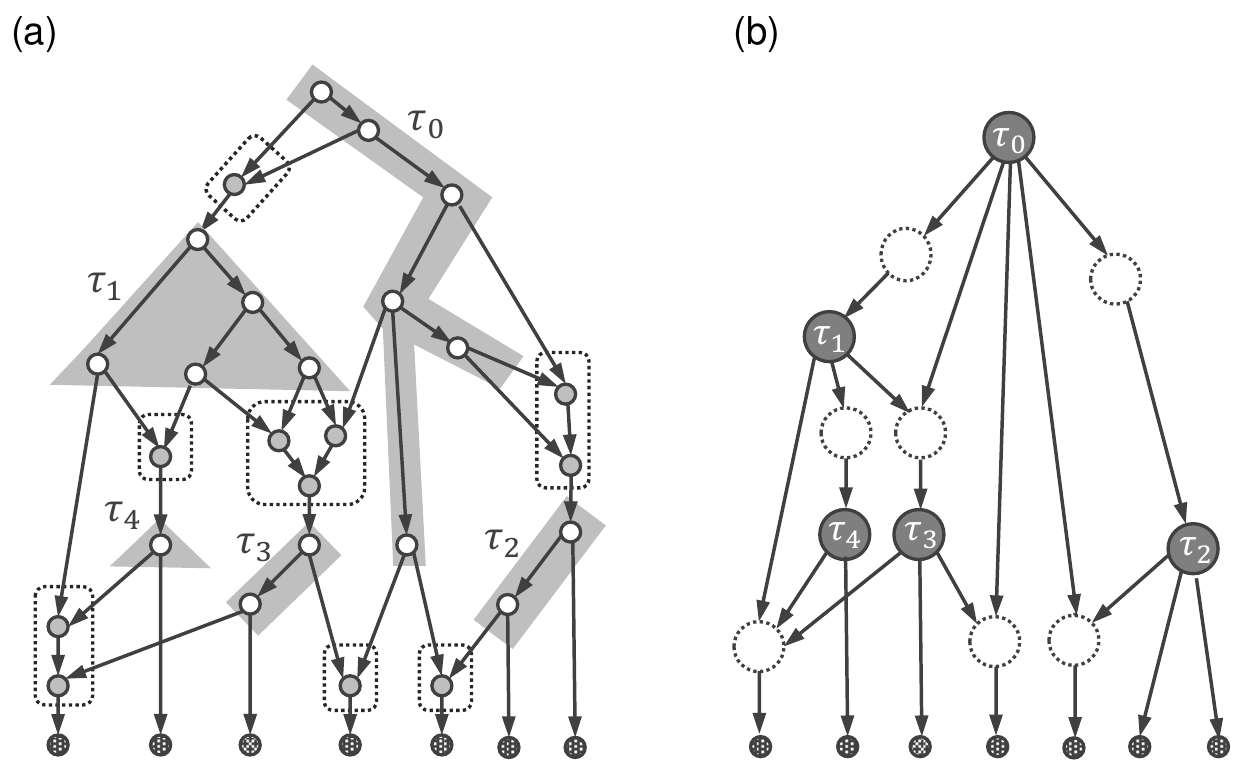}
            \caption{Illustration of the concept of the compression. ({\bf a}) An RPN with five tree-node components $\tau_0$--$\tau_4$ (shaded) and seven reticulation components (dashed rectangles).  ({\bf b}) The compression of the  RPN in Part a. Here, each node represents either a tree node component (filled) or a  reticulation component (dashed) in the original network. Note that the three degree-2 nodes represent reticulation components in the original network. }
            \label{Fig_1}
        \end{figure}

Consider an RPN $N$ over a set of taxa. Recall that 
$\mathcal{D}(N)$ and $\mathcal{L}(N)$ denote the sets of degree-2 nodes and leaves of $N$, respectively.  Assume that $N$ contains $p+1$ tree-node components:
 $$T_N=\{\tau_0, \tau_1, \cdots, \tau_p\}$$ and 
$q$ reticulation components: 
$$\Sigma_N=\{\sigma_1, \sigma_2, \cdots, \sigma_q\}.$$ The compression of $N$ is an RPN $\bar{N}$ over the same taxon set defined by:
\begin{eqnarray}
 & &\mathcal{V}(\bar{N})=T_N \cup \Sigma_N 
  \cup \mathcal{D}(N) \cup \mathcal{L}(N), \\
&& \mathcal{E}(\bar{N})=E_1 \cup E_2 \cup E'_2 \cup E_3, 
\end{eqnarray}
where
\begin{eqnarray*}
&&E_1=\{  (u, v)\in \mathcal{E}(N) \;|\; u\in \mathcal{D}(N) \;\& \;
    v\in \mathcal{D}(N) \cup \mathcal{L}(N)\},\\
&&E_2=\{ (x, v) \;|\; u\in x\in (T_N\cup \Sigma_N)  \;\&\; v\in (\mathcal{D}(N) \cup \mathcal{L}(N)) \;\&\;  (u, v)\in \mathcal{E}(N) \},\\
&&E'_2=\{ (v, y) \;|\; u\in y\in (T_N\cup \Sigma_N)  \;\&\; v\in \mathcal{D}(N) \;\&\;  (u, v)\in \mathcal{E}(N) \},\\
&& E_3=\{ (x, y) \;|\; u\in x\in (T_N\cup \Sigma_N)  \;\&\; v\in y\in (T_N\cup \Sigma_N) \;\&\;  (u, v)\in \mathcal{E}(N) \}.
\end{eqnarray*}
Here, that $w\in x$ means that $w$ is a tree (resp. reticulate) node and $x$ is a tree-node (resp. reticulation) component such that $w$ is a node in $x$. 
The definition of the compression is illustrated in Fig.~\ref{Fig_1}, where the network does not contain any degree-2 nodes.

For a tree  (resp. reticulate) node $u$, we use $\tau _u$ (resp. $\sigma_u$) to denote the tree-node (resp. reticulation) component that contains $u$.  We define a surjective mapping $f:  \mathcal{V}(N) \rightarrow \mathcal{V}(\bar{N})$ by:
\begin{eqnarray}
\label{def_f}
   f(u)=\left\{\begin{array}{cl}
            \tau_u, & u\in \mathcal{T}(N),\\
            \sigma_v, & u\in \mathcal{R}(N),\\
            u, & u\in (\mathcal{D}(N)\cup \mathcal{L}(N))
          \end{array} \right.
\end{eqnarray}
and a partial mapping $g: \mathcal{E}(N) \pto \mathcal{E}(\bar{N})$ by:
\begin{eqnarray}
  g(e)=(f(u), f(v))
\end{eqnarray}
for any $e=(u, v)$ such that  $\{u, v\} \not\subseteq \mathcal{T}(N)$ and  $\{u, v\}\not\subseteq  \mathcal{R}(N)$. 
It is easy to verify that $g$ is surjective.

We further define:
\begin{eqnarray}
&& f^{-}(a)=\{u \in \mathcal{V}(N) \;|\; f(u)=a\}, \mbox{ for }  a \in \mathcal{V}(\bar{N}),\\
&&f(X)=\{f(x) \;|\; x\in X\}, \mbox{ for }  X \subseteq \mathcal{V}(N),\\
 &&g(Y)=\{ g(e) \;|\; e\in Y  \;\&\;  \mbox{$g$ is defined on $e$}\}, \mbox{ for } Y\subseteq \mathcal{E}(N).
\end{eqnarray}

\begin{proposition} 
\label{prop_3.2}
Let $M$ be a sub-RPN of $N$ and $\bar{M}$ be $(f(\mathcal{V}(M)), g(\mathcal{E}(M)))$.
\begin{itemize}
 \item[{\rm (1)}]  $\bar{M}$ is also a directed path or a node if $M$ is a directed path.
\item[{\rm (2)}] $\bar{M}$ is connected if $M$ is connected.
\item[{\rm (3)}] $\bar{M}$ is a tree if $M$ is a tree.
\end{itemize}
\end{proposition}
{\bf Proof.}  (1) Assume $M$ is a directed path $u_0, u_1, \cdots, u_k$ ($k\geq 1$). Then, for each $i\leq k-1$, 
  by definition, $f(u_i)=f(u_{i+1})$ or $(f(u_i), f(u_{i+1}))\in \mathcal{E}(\bar{N})$. Hence, removing identical  nodes from $f(u_0), f(u_1), \cdots, f(u_k)$ gives a node or a directed path in $\bar{N}$.

 (2) and (3)  follow from Part 1. \QED

\begin{proposition}
The compression of any tree-based RPN is also tree-based.
\end{proposition}
{\bf Proof.} Assume $N$ is tree-based. Then, there exists a subset $E$ of reticulate edges such that 
  such that $T=(\mathcal{V}(N), \mathcal{E}(N)\backslash E)$ is a subtree such that its leaf set is exactly equal to $\mathcal{L}(N)$. For each tree-node component $\tau$ of $N$,  there is only one edge $(u, v)$ in $N$ such that $u\not \in \tau \ni v$.  This edge must be in the subtree $T$. 

For each reticulation component $\sigma$,  its nodes induce a path 
in $T$. This implies that $T$ contains only one edge $(u, v)$ such that $u\not\in \sigma \ni v$. 

Taken together, by Proposition~\ref{prop_3.2}, these two facts indicate that each non-root node is of indegree 1 in the subgraph $\bar{T}=(\mathcal{V}(\bar{N}), g(\mathcal{E}(N) \backslash E))$, implying that $\bar{T}$ is a subtree with the leaf set
$\mathcal{L}(N)$ in $\bar{N}$, as only the leaves of $T$ are mapped to the leaves of $\bar{T}$.   
\QED

 A binary RPN is {\it galled} if every reticulate node $u$ has an ancestor $w$ such that $w$ is a tree node and there are two
internally disjoint paths from $w$ to $u$ in which all nodes but $u$ are each a tree node.

\begin{theorem}
\label{main_thm}
Let $N$ be an RPN over a taxon set. Then, 
 \begin{itemize}
  \item[{\rm (1)}] $\bar{N}$ is  a tree  if $N$ is binary and galled.
  \item[{\rm (2)}] $\bar{N}$ is a  tree-child network if $N$ is reticulation-visible. 
 \end{itemize} 
\end{theorem}
{\bf Proof.}  
Galled networks are reticulation-visible \cite{Huson_Book_11}. Let $N$ be a reticulation-visible network.  Let $u$ be a reticulate node of $N$. Since $u$ is visible,  by Proposition~\ref{prop_1}, its unique child is a visible tree node. Thus, every reticulation component contains only one reticulate node in $N$. 

(1).  In an RPN, a reticulate node $v$ is {\it inner} if all its parents belong to the same tree-node component of the network.
If $N$ is galled, then every reticulate node is inner \cite{Gunawan_IC_17}. For any reticulation component $\sigma =\{v\}$, we use $\tau$ to denote the tree-node component that contains all the parent of $v$ in $N$. By definition, $\tau$ is the only parent of $\sigma$ in $\bar{N}$. Therefore, 
$\sigma$ is of indegree 1 and outdegree 1. This implies that $\bar{N}$ does not contain any reticulate node and every node is of indegree 1. Thus, $\bar{N}$ is a tree. 

(2). Let $N$ be reticulation visible and  $\tau$ be a tree-node component of $N$. 
Since $N$ is reticulation-visible, one of the following three conditions holds: (a) $\tau$ contains  the parent of some  $\ell$; (b) $\tau $ contains the parent of some redundant node $d$; and (c) $\tau$ contains all the parents of a reticulate node $v$ (see \cite{Gunawan_IC_17}). If  (a) (resp. (b))  holds, 
$\ell$ (resp. $d$) is the child of $\tau$ in $\bar{N}$. If Condition c holds, the node $f(\{v\})$ is of degree-2 and is the child of $\tau$ in $\bar{N}$.   

Let $x$ be a redundant or reticulate node in $N$. Since it is visible, its child $c(d)$ must be a tree or redundant node and so $f(d)$ has $f(c(d))$ as a child. 
Thus, every non-leaf node has a child that is a tree or redundant node, implying that $\bar{N}$ is tree-child.  
\QED

Lastly, we point out that the class of tree-sibling RPNs is not closed under compression. For example,  the binary RPN in 
Fig.~\ref{Fig_0}c is tree-sibling. Compressing it results in the RPN in Fig.~\ref{Fig_0}d that is not tree-sibling.


\section{An Application of Theorem~\ref{main_thm}}
\label{sec4}

\subsection{New network classes}

In group theory, group homomorphism and quotient group are important concepts. These concepts enable us to understand the structures of abelian groups \cite[page 176]{Rose_Book_12}. 
We introduce network compression with the same spirit.  It enables us to examine reticulation-visible networks from a different angle. For example, 
Theorem~\ref{main_thm}  suggests that reticulation-visible networks are networks that are expanded from tree-child network by replacing some nodes with trees. 
 Meanwhile, the theorem can  also be used to define new classes of RPNs. For instance, 
Fig.~\ref{Fig_1} presents a binary RPN that has a tree-child compression,  but it is not reticulation-visible. 

\begin{definition}
 An RPN $N$ is quasi-reticulation-visible (resp. quasi-galled) if and only if $\bar{N}$ is a tree-child (resp. tree) network. 
\end{definition}

 \begin{figure}[b!]
            \centering
            \includegraphics[scale = 1.0]{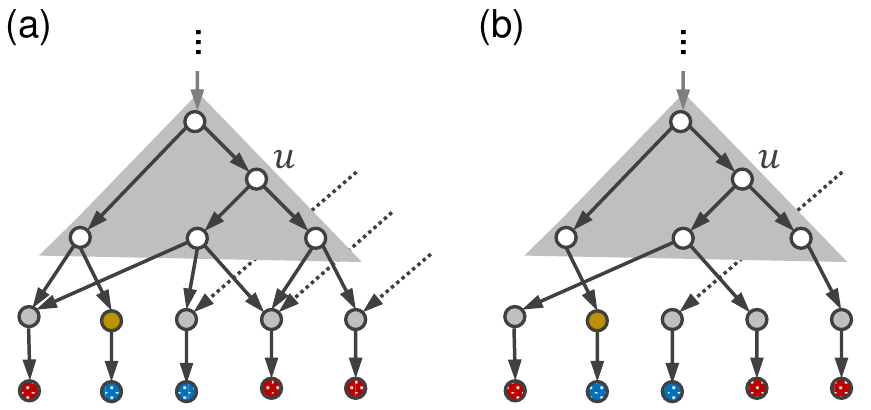}
            \caption{Illustration of Proposition~\ref{prop_4.1}, where $N$, $u$ and $S$ are defined.  ({\bf a}) An exposed tree-node  component  $\tau$ of $N$,  below which there are three leaves (red) that form $S$ and two other leaves (blue), one redundant node (orange) and four isolated reticulate nodes. Dashed edges are the reticulate edges from nodes not in $\tau$. ({\bf b}) For each leaf $\ell$, we remove all edges entering $p(\ell)$ except for one  from a node below $u$ in $\tau$ if $\ell$ is red and not below $u$ (which may or may not in $\tau$) if $\ell$ is blue, where $p(\ell)$ is the reticulate parent of $\ell$. This cutting process  results in a subtree rooted at $u$ that contains all red leaves.}
            \label{Fig3_exposed}
        \end{figure}

\subsection{Algorithm Design}

The connections between clusters, trees and RPNs  are basic issues in the theoretical study of phylogenetic networks. For example,   different distance metrics based on clusters and trees are defined  in the space of RPNs \cite{Huson_Book_11}. Hence,  the cluster containment problem has been studied.

In a phylogenetic tree, the subset of labeled leaves below a node  is called the {\it cluster} displayed at the node. 
 Consider an RPN $N$ over $X$  and $u\in \mathcal{T}(N)$. A subset $S\subseteq X$ is displayed as a softwired cluster  at $u$ if 
there is a spanning tree $T$ of $N$ in which $S$ is displayed at $u$, where some leaves of $T$ are unlabeled non-leaf nodes of  $N$ and not counted when computing a softwired cluster. 
\vspace{0.5em} 

\noindent {\bf Small Cluster Containment} (SCC)\\
\noindent {\sc Instance}: An RPN over $X$, $S\subseteq X$ and a tree node $u\in \mathcal{T}(N)$.\\
\noindent {\sc Question}: Is $S$ displayed at $u$ in $N$?
\vspace{0.5em}

\noindent {\bf Cluster Containment} (CC)\\
\noindent {\sc Instance}: An RPN over $X$ and $S\subseteq X$.\\
\noindent {\sc Question}: Is $S$ displayed at some node in $N$?
\vspace{0.5em}

A reticulate node is {\it isolated}  if neither its parents nor child is reticulate.
A tree-node component $\tau$ is  {\it exposed}  if only leaves, isolated reticulate nodes and redundant nodes are found below it (Fig.~\ref{Fig3_exposed}a).

\begin{proposition}
\label{prop_4.1}
  Let $N$ be an RPN over $X$ and $u$ be a tree node in an exposed tree-node component $\tau$. Then, for any $S\subseteq X$,  $S$ is displayed at $u$ if and only if the following two statements hold:
\begin{itemize}
 \item[{\rm (i)}] Every leaf $\ell \in S$ is below $u$, and
 \item[{\rm (ii)}] Every leaf $\ell \not\in S$ is not dominated by $u$.
\end{itemize} 
\end{proposition}
{\bf Proof.} Assume that $S$ is displayed at $u$ in $N$. Then, $N$ has a spanning tree $G$ in which $S$ is the cluster of $u$. Since each leaf $\ell \in S$ is below $u$ in $G$, it is below $u$ in $N$, implying Fact i.  Since every leaf $\ell\not\in S$ is not found below $u$ in $G$,  the path from the root to $\ell$ in $G$ does not contain $u$, implying that $\ell$ is not dominated by $u$ in $N$.

Conversely, assume that Facts i and ii are true.  Since non-reticulate nodes below $\tau$ are  leaves and redundant nodes only and reticulate node below $\tau$ are isolated,  there is exactly one leaf below each reticulate node below $\tau$. For a reticulate node $x$ below $\tau$, we let $\ell_x$ denote its unique  leaf child below $x$ and consider the following cases separately. 

If $\ell_x$ is in $S$, it is below $u$. Hence, there is a directed path $P(u, x)$ from $u$ to $x$ and then to $\ell_x$.  We then remove all reticulate edges entering $x$ that are not in $P(u, x)$.  

If $\ell_x$ is not in $S$, the fact that $\ell_x$ is not dominated by $u$ implies that there is a path from the root of $N$ to $x$ that does not contain $u$.  In this case, we also remove all reticulate edges entering $x$ that are not in the path. 

 After repeating the above cutting process for every reticulate node below $\tau$, the subgraph below $u$ in the resulting network is a subtree containing all leaves of $S$, implying that $S$ is a cluster of $u$ in $N$.
\QED

In the rest of this section, we study the SCC problem on RPNs that are {\bf quasi-reticulation-visible}. Let $N$ be a quasi-reticulation-visible RPN on $X$. Without loss of generality, we may assume that $N$ does not contain any degree-2 node.
By definition, $\bar{N}$ is a tree-child network in which  reticulate nodes are isolated and each non-leaf node is connected to some leaf by a path consisting of tree nodes and redundant nodes only.  Here, we emphasize that redundant nodes are not considered to be reticulate in $\bar{N}$, although they correspond one-to-one with the reticulation components $\sigma$ of $N$ with the property that 
$f(w)=f(w')$ for any two edges $(w, v)$ and $(w', v')$ such that $f(v)=f(v')=\sigma$, where $f(\;)$ is defined in 
Eqn.~(\ref{def_f}).

\begin{figure}[h!]
            \centering
            \includegraphics[scale = 0.8]{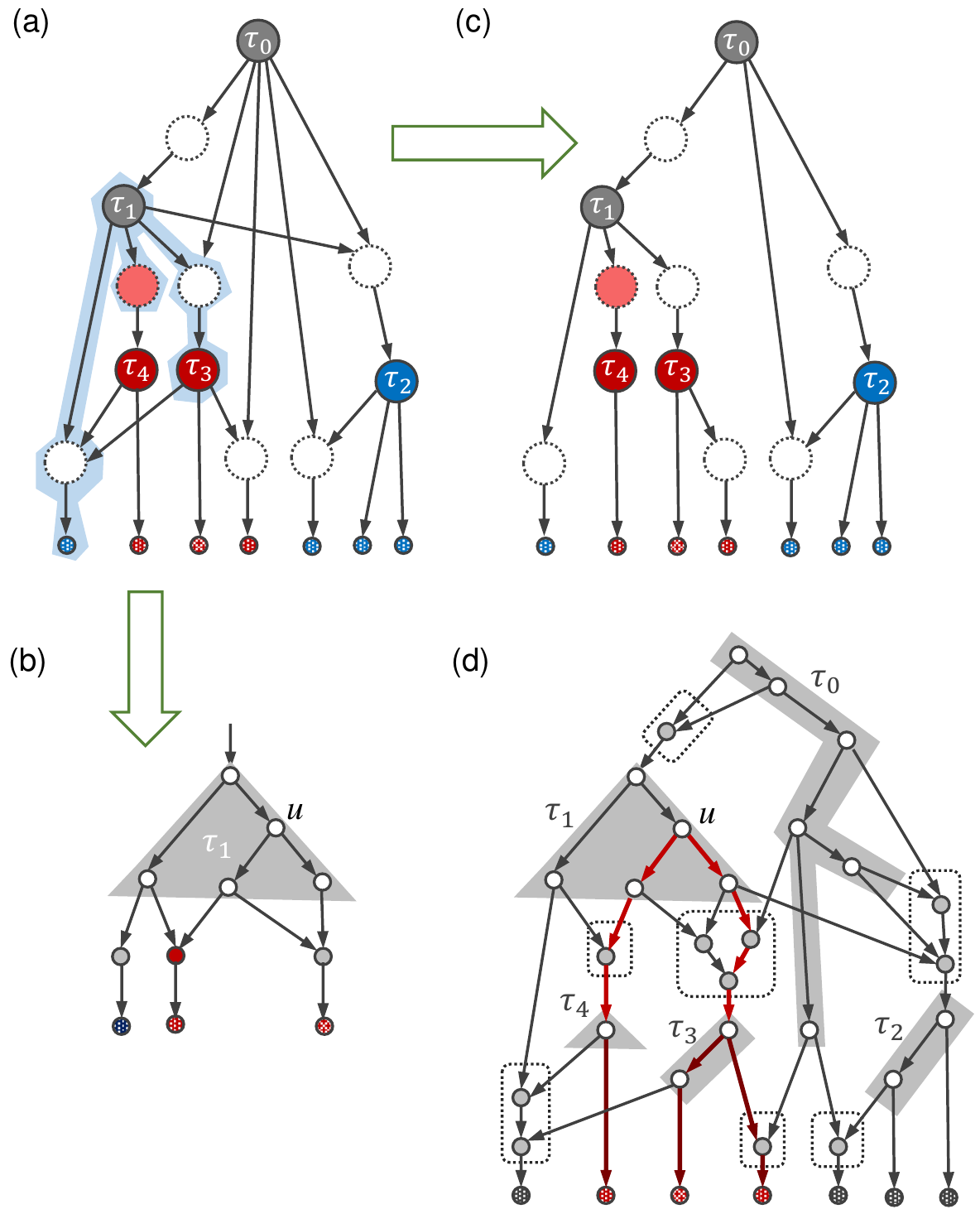}
            \caption{Illustration of Proposition~\ref{prop_4.2}. A quasi-reticulation-visible network $N$, a tree node $u$ and a set of leaves (red) in Panel {d}.   ({\bf a}) The compression $\bar{N}$ of $N$ in which non-reticulate nodes below $f(u)$ are colored. 
 ({\bf b}) The network $N'$ defined using the tree-node component $\tau_u$ (which is $\tau_1$) that contains $u$ and all nodes around $f(u)$ in $\bar{N}$.
 ({\bf c}) The network $N''$ defined by removing edges from red (resp. blue) nodes to reticulate nodes with blue (resp. red) children. ({\bf d})  The red subtree shows that the given subset of leaves (red) is displayed at $u$. The top (red) and bottom (dark red) parts of this subtree are derived from $N'$ and the subnetwork rooted at $f(u)$ of $N''$, respectively.}
            \label{Fig4_condition}
 \end{figure}

Furthermore,  for determining whether $S$ is a softwired cluster at $u$ in $N$,  we just need to distinguish the leaves in $S$ from the others.
Hence, we color the leaves of $S$ red and the others blue. 
We further extend the leaf coloring to color non-reticulate nodes that are below  $f(u)$ in $\bar{N}$. For each non-reticulate node 
$w$ that is below $f(u)$,  it is colored: 
\begin{itemize}
 \item  Red (resp. blue) if  all the non-reticulate children of $w$ are red (resp. blue). 
 \item  Purple if $w$ has either at least one purple child or at least one blue child and one red child.
\end{itemize}
Since $\bar{N}$ is tree-child, each non-reticulate node below $f(u)$ will be assigned a color through this bottom-up coloring extension (Fig.~\ref{Fig4_condition}a). It is easy to see that this coloring extension can be done in linear time. 
Note that  {\bf $f(u)$ and all the nodes not below $f(u)$ have not been colored}.

For each $w\in \mathcal{T}(N)$,  we use:
\begin{itemize}
\item $\tau_w$ to denote the tree-node component containing $w$ in $N$,  and 
\item $f(w)$ to denote the node in $\bar{N}$ that represents $\tau_w$.
\end{itemize} 
Next, using the partially colored $\bar{N}$, we define a new network $N'$ from  $\tau_u$, in which $\tau_u$  is  exposed and there is no other tree-node component. 
The node set of $N'$ is the union of the following node subsets:
\begin{itemize}
   \item[1.] The tree nodes within  $\tau_u$;
   \item[2.] Every leaf that is a child of $f(u)$ in $\bar{N}$; 
   \item[3.] Every redundant node $x$ that is a child of $f(u)$ in $\bar{N}$ and a new leaf $\ell_x$ with the same color as $x$;   
   \item[4.] Every reticulate child $y$ of $f(u)$ in $\bar{N}$ such that  (i) $y$ has a blue child and (ii) the parents of $y$ are all red except $f(u)$, and a new leaf  $\ell_y$ of blue;  
\item[5.]  Every reticulate child $z$ of  $f(u)$ in $\bar{N}$ such that (i) $z$ has a red child and (ii) $z$ does not have any red parents, and a new leaf $\ell_z$ of red.
\end{itemize}
The edge set of $N'$ is the union of the following edge subsets:
\begin{itemize}
\item The edges within $\tau_u$; 
\item The reticulate edge set $\{(v, x) \;|\; (v, w)\in \mathcal{E}(N) \mbox{ for }  v\in \tau_u, w\in \sigma_x\}$, where $x$ is a node added in Items 3\textendash 5 and $\sigma_x$ is the reticulation component corresponding with $x$ in $N$; 
\item The edges $(v, \ell)\in \mathcal{E}(N)$ for $\ell \in \mathcal{L}(N)$ appearing in Item 2, where $v$ is in $\tau_u$. 
\item The edges $(x, \ell_x)$ so that $\ell_x$ is the child of $x$ for all $x$ and $\ell_x$ are defined in Items 3--5. 
\end{itemize}
How to construct $N'$ is  illustrated in Fig.~\ref{Fig4_condition}.  For this example, the tree-node component 
containing $u$ is $\tau_1$ and so $f(u)=\tau_1$ in $\bar{N}$ (Fig.~\ref{Fig4_condition}a).  In $\bar{N}$, $\tau_1$
has no leaf child,  one redundant child and three reticulate children. The right-most reticulate child of $\tau_1$ that has a blue child has $\tau_0$ as an uncolored parent and hence does not appear in $N'$ (Fig.~\ref{Fig4_condition}b).  The other three reticulate children of $\tau_1$  (shaded, Fig.~\ref{Fig4_condition}a) appear  in  $N'$. 

Lastly, we obtain  a subnetwork $N''$ of $\bar{N}$ by applying the following process:
For each reticulate node $x$ that is below $f(u)$,  let $c(x)$ be its child in $\bar{N}$.
\begin{itemize}
 \item  If $c(x)$ is red, remove all edges  entering $x$ from blue and uncolored parents other than $f(u)$. Additionally, if $x$ has a red parent, remove also the edge from $f(u)$ if exists.
 \item  If $c(x)$ is blue, remove all incoming edges  entering $x$ from red parents. Additionally,  if $x$ has a blue or uncolored parent other than $f(u)$, remove also the edge from $f(u)$ if exists.
\end{itemize} 
The construction of $N''$ is illustrated in Fig.~\ref{Fig4_condition}. There are five reticulate nodes below $f(u)$ in $\bar{N}$. The left-most one has a blue leaf as a child and has two red parents  $\tau_3$ and $\tau_4$. Hence, the edges from $\tau_3$ and $\tau_4$ to this reticulate node were removed (Fig.~\ref{Fig4_condition}c). Similarly, we removed the edges from $\tau_1$ to the parent of $\tau_2$,  from $\tau_0$ to the parent of $\tau_3$ and from 
$\tau_0$ to the parent of the fourth leaf, which  is red.

\begin{proposition}
\label{prop_4.2}
 Let $N$ be a quasi-reticulation-visible RPN over $X$, $X\subseteq S$ and $u\in \mathcal{T}(N)$.
Then, $S$ is displayed at $u$ in $N$ if and only if the following three statements hold:
\begin{itemize}
\item[{\rm (i)}]   No  purple node is below $f(u)$ in $\bar{N}$.  
\item[{\rm (ii)}]  The set of red leaves is displayed at $u$ in $N'$.
\item[{\rm (iii)}]  $N ''$ is connected in which all the red leaves are below $\tau_u$.
\end{itemize}
\end{proposition}
{\bf Proof.}  Assume that $S$ is displayed at $u$ in $N$. Then, $N$ has a spanning tree $G$ in which  the red leaves are exactly those below $u$. Let $\bar{G}= ( f(\mathcal{V}(G)), g(\mathcal{E}(G)))$ and $\overline{G(u)}=( f(\mathcal{V}(G(u))), g(\mathcal{E}(G(u))))$, where $G(u)$ is the subtree of $G$ rooted at $u$.  By Proposition~\ref{prop_3.2}, $\bar{G}$ is a spanning tree of $\bar{N}$ and  $\overline{G(u)}$ is a subtree of 
$\bar{G}$  that contains exactly all the red  leaves. 
For example, if $G(u)$ is the spanning tree highlighted in red in Fig.~\ref{Fig4_condition}d, 
$\overline{G(u)}$ consists of the nodes in the paths from 
$\tau_1$ to the three red leaves (Fig.~\ref{Fig4_condition}c).  Note that $\overline{G(u)}$ is not the subtree rooted at $f(u)$ of $\bar{G}$ in general.

If there is a
purple tree node $v$ below $f(u)$,  a red leaf $\ell_r$ and a blue leaf $\ell_b$ exist such that there are paths $P'$ and $P''$ from $v$ to $\ell_r$ and $\ell_b$, respectively,  that consist of non-reticulate nodes only, according to how tree nodes are colored. Clearly, $P'$ and $P''$ are also in 
$\overline{G(u)}$, contradicting that only red leaves are found below $u$ in $\overline{G(u)}$.  This completes the proof of Fact i. 

Consider a  red leaf $\ell'$  in $N'$.   It represents a red leaf $\ell$ of $N$ or corresponds with a red node $x$ of $\bar{N}$ that represents a tree-node component whose root is a dominator of a red leaf $\ell_x$  in $N$.  
If the former is true, then, there is path from $u$ to $\ell$.  If the latter is true,  there is a path from $u$ to $\ell_x$ that contains the root of $x$ in $G$. Taken together, the two facts imply that $\ell'$ is below $u$ in $N'$.

Consider a blue leaf $\ell'$ in $N'$.  It represents a blue leaf $\ell$ of $N$ or corresponds with a blue node $y$ of $\bar{N}$ that represents a tree-node component whose root is a dominator of a blue leaf $\ell_y$  in $N$.
If the former is true, there is path from the root of $G$ to $\ell$  that avoids $u$. If
the latter is true,  there is a path from the network root  to $\ell_y$ that contains the root of $y$ and avoids $u$ in $G$. This implies that $\ell'$ is not dominated by $u$ in $N'$. 

Taken together,  by Proposition~\ref{prop_4.1},  the facts in the two paragraphs above imply that Fact ii is true. 

Since $\bar{N}$ is tree-child and $\bar{G}$ is a spanning tree of $\bar{N}$,  $\bar{G}$ contains every path consisting of non-reticulate nodes in  $\bar{N}$.
Therefore,  for each reticulate node below $f(u)$, it connects a red node with either a red node or $f(u)$,  or it connects a blue node with a blue or uncolored node in   $\bar{G}$.  This implies  that $\bar{G}$ is subtree of $N''$. Therefore, $N''$ is connected and all red leaves are below $f(u)$ in $N''$, implying that Fact iii is true.

Conversely, assume that the three statements hold. By Fact i, no purple node is found in $\bar{N}$ and hence $N''$. 
By Fact iii, all red leaves are below $f(u)$ in $N''$.  For each reticulate node having a red (resp. blue) child, by constrcution,  
$f(u)$ is its only parent,  or else its parents are all red (resp. blue or uncolored) in $N''$.  Thus, any spanning tree $G_u$ of the subnetwork rooted at $f(u)$ of $N''$ has the following property:
\begin{quote}
  In $G_u$, the path from $f(u)$ to $\ell$  consists of red (resp. blue) nodes (excluding $f(u)$) and reticulate nodes only,
  for each red (resp. blue) leaf $\ell$ below $f(u)$.
\end{quote}
This implies that a subset of red leaves is displayed at each red grandchild of $f(u)$ by $G_u$ such that their union contains all the red leaves.
 Note that  these subsets of red leaves  are also displayed at the roots of the corresponding tree-node components in $N$ simultaneously. 

Most importantly, by construction, each leaf  of $N'$ corresponds  with a unique  node that is either a leaf child of $f(u)$ or a grandchild of  $f(u)$ in $N''$. 
Taken together, these facts and Fact ii imply that the set of all red leaves are displayed at $u$ in $N$, as showed in  Fig.~\ref{Fig4_condition}d.  \QED

Since the three statements in Proposition~\ref{prop_4.2} can be checked in linear time, we obtain the following resutls. 
\begin{corollary}
There is a linear-time SCC algorithm for quasi-reticulation-visible networks. 
\end{corollary} 

\noindent {\bf Remarks}  The approach presented in this section can be modified to design linear-time algorithms for solving the CC problem and the tree containment problem.

\section{Conclusion}
\label{sec5}

We have formally introduced the concept of compression for RPNs.  Using it, we have presented another interesting connection between reticulation-visible network and tree-child network and introduced a new network class for which the CC problem is solvable in linear-time.  We have also showed that tree-sibling RPNs are not closed under compression, an undesired property of this network class.  

We introduce network compression in analogy with quotient group in group theory.  The fundamental (or basis) theorem of finite abelian groups states that every finite abelian group can be expressed as the direct product (or sum) of cyclic subgroups of prime-power order \cite{Rose_Book_12}. 
 Is network compression  useful for network reconstruction?  This is definitely worth studying in future.

\end{document}